\documentclass[runningheads]{llncs}
\usepackage[T1]{fontenc}

\usepackage{cite}
\usepackage{times}
\usepackage{multirow}
\usepackage{graphicx}
\usepackage{multicol}
\usepackage{algorithm}
\usepackage{algpseudocode}
\usepackage{subcaption}
\usepackage{booktabs}

\usepackage{hyperref}
\usepackage{url}
\usepackage{xcolor}
\usepackage{tcolorbox}
\usepackage{graphbox}
\usepackage{tikz}
\usepackage{pifont}
\usepackage{colortbl} 
\usepackage{xspace}
\usepackage{makecell}
\usepackage{arydshln}

\hypersetup{
    colorlinks,
    linkcolor={blue!80!black},
    citecolor={blue!80!black},
    urlcolor={blue!80!black}
}
\definecolor{lightyellow}{RGB}{255, 255, 190} 
\newcommand{\roundedcell}[2]{%
  \tikz[baseline=(char.base)]\node[draw=none, fill=#1, rounded corners=2pt, inner sep=2pt] (char) {#2};%
}
\newcommand{\method}{UNDREAM\xspace}

\usepackage{amsmath,amsfonts,bm}
\usepackage{booktabs}
\usepackage{graphicx}

\begin{document}
\title{Differentiable Rendering Powered\\ End-to-End Adversarial Attack Evaluation}

\titlerunning{Differentiable Rendering Powered End-to-End Adversarial Attack Evaluation}
\author{Mansi Phute\inst{1}\orcidID{0000-0003-4682-6281} \and
Matthew Hull\inst{1}\orcidID{0000-0002-3451-0901} \and
Haoran Wang\inst{1} \orcidID{0009-0009-6005-4952}\and
Alec Helbling\inst{1} \orcidID{0009-0007-8846-6460}\and
ShengYun Peng\inst{1}\orcidID{0000-0003-3063-2052} \and 
Willian Lunardi\inst{2} \orcidID{0000-0003-0718-0019}\and 
Martin Andreoni\inst{2} \orcidID{0000-0002-4170-4341}\and 
Wenke Lee\inst{1} \orcidID{0000-0003-2761-1277}\and 
Duen Horng Chau\inst{1} \orcidID{0000-0001-9824-3323}
}
\authorrunning{M. Phute et al.}
\institute{Georgia Institute of Technology, USA \and
Technology Innovation Institute, UAE}
\maketitle              %
\begin{figure}[h]
    \centering
    \includegraphics[width=\textwidth]{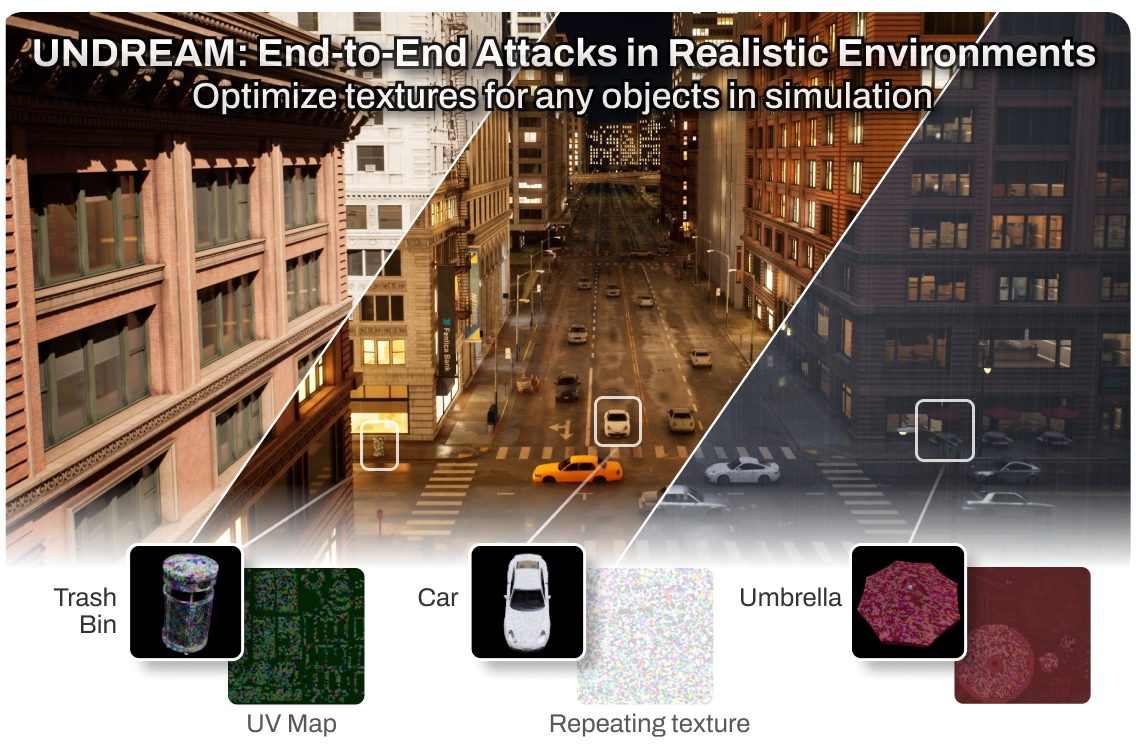} 
    \caption{\method is the first software framework bridging differentiable rendering and photorealistic simulation to enable end-to-end adversarial attacks. Users can create diverse environments by controlling environmental conditions, add and configure custom 3D objects and execute adversarial attacks that faithfully follow threat model.}
    \label{fig:crown-jewel}
\end{figure}

\begin{abstract}
Deep learning models deployed in safety critical applications like autonomous driving use simulations to test their robustness against adversarial attacks in realistic conditions. 
However, these simulations are non-differentiable, forcing researchers to create attacks that do not integrate simulation environmental factors, reducing attack success. 
To address this limitation, we introduce \method, the first software framework that bridges the gap between photorealistic simulators and differentiable renderers to enable end-to-end optimization of adversarial perturbations on any 3D objects.
\method enables manipulation of the environment by offering complete control over weather, lighting, backgrounds, camera angles, trajectories, and realistic human and object movements, thereby allowing the creation of diverse scenes.
We showcase a wide array of distinct physically plausible adversarial objects that \method enables researchers to swiftly explore in different configurable environments. 
This combination of photorealistic simulation and differentiable optimization opens new avenues for advancing research of physical adversarial attacks.

\end{abstract}

\section{Introduction}

\begin{figure}[h]
    \centering
    \includegraphics[width=
    \textwidth]{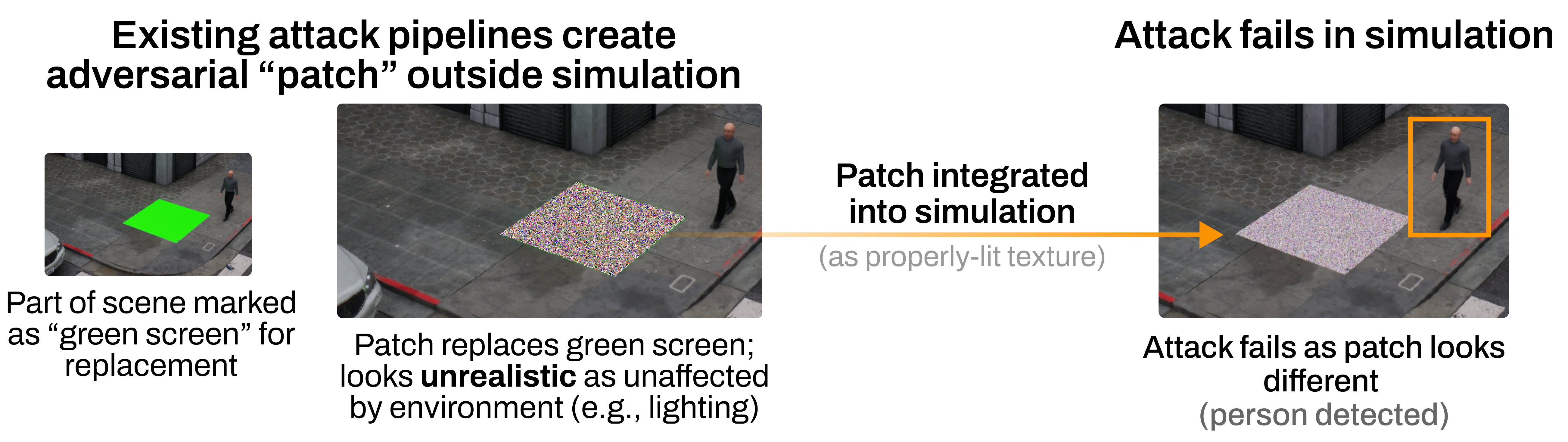} 
    \caption{Existing attack pipelines create adversarial attacks outside of simulation without accounting for interaction with the environment, resulting in attacks that fail when the patches are integrated in the simulation.}
    \label{fig:reinsert}
\end{figure}

Ensuring the adversarial robustness of vision systems is important, as computer vision is applied in safety-critical domains like autonomous vehicles. For example, adversarial perturbations when put on stop signs can deceive object detection systems, potentially causing them to misclassify or fail to recognize traffic signs altogether \cite{wei2024physical}. 
Adversarial perturbations are typically generated through an iterative optimization process without ever inserting the perturbations into the simulator itself. 
This begins with an initial pattern, commonly referred to as a ``\textit{patch},'' alongside a collection of environmental images in which the patch is to be evaluated. 
This patch is optimized for the desired adversarial task. 
The optimization process does not account for complex interactions between the patch and the environment, such as lighting variations, or material properties. 
As a result, when the optimized patch is eventually placed into the simulation environment, its visual appearance often changes dramatically, due to these unmodeled physical factors, leading to reduced adversarial effectiveness, as shown in \autoref{fig:reinsert}. 
This observation aligns with findings from prior work, which reports similar degradation in performance under such conditions \cite{xu2024imperceptible}. 
This issue is particularly concerning, given that simulations are intended as a proxy for real- world situations \cite{wei2024physical}. 
Consequently, the failure of the attack to remain effective under realistic lighting and material conditions raises significant concerns about its transferability from digital to physical environments \cite{nesti2022evaluating, hu2023physically}.

\noindent
\textbf{Why do researchers continue to use this sub-optimal pipeline?} 
A primary reason is that the photorealistic simulation platforms such as Unreal Engine \cite{engine2018unreal} that allow for complex human and object movements, and realistic physics modules for collision and light scattering \cite{wei2024physical}, are \textbf{all non-differentiable} (\autoref{tab:comparison}).
Researchers have tried inserting patches into the simulation itself and estimating gradients using PyTorch transformations \cite{xu2024imperceptible}. 
However, such approaches restrict the patch to shapes and transformations that can be approximated in two dimensions,
limiting their applicability in creating complex 3D adversarial objects \cite{nesti2024carla}. 
While differentiable renderers are often used in the creation of adversarial textures they lack other attributes to act as a satisfactory substitute, such as simulating movement of objects and an easy-to-use interface \cite{hull2024renderbender}. 
This fundamental gap explains why recent adversarial methods use this optimization pipeline or other similar approximation techniques; direct access to gradients within realistic simulation environments is unavailable, preventing end-to-end optimization of textures
\cite{xu2024imperceptible}. %
As a result, there is not a unified approach that enables differentiable optimization within 3D simulation, a critical gap that creates a barrier to research progress \cite{wei2024physical}.

We propose \textbf{\method}: \textbf{U}nified  \textbf{N}atural \textbf{D}ifferentiable \textbf{R}endering for \textbf{E}nd-to-end \textbf{A}dversarial \textbf{M}odeling, a novel practical  framework that bridges photorealistic 3D simulation and differentiable optimization, offering the best of both worlds. \method converts 3D objects and textures from Unreal Engine into a fully differentiable system, enabling end-to-end optimization of adversarial patterns on arbitrarily-shaped 3D objects within photorealistic environments. We make the following key contributions:
\begin{enumerate}
\item \textbf{\method: First Software Framework Bridging Differentiable Rendering and Photorealistic Simulation for End-to-End Adversarial Attacks.} 
Unlike existing approaches that optimize textures in isolation and later superimpose them (\autoref{fig:reinsert}), \method optimizes textures as they appear natively in simulation, preserving lighting, perspective, and physical material interactions. (\autoref{sec:attack-pipeline}) %

\item \textbf{Automatic, Precise 3D Transformation Enabling Faithful Threat Modeling.} 
By embedding adversarial textures directly into the simulation environment, \method \textbf{eliminates} the need for 
calculating the bounds of the adversarial object, 
leading to better alignment with realistic threat models and provides a more faithful evaluation of adversarial attacks. (\autoref{sec:inverse-rendering}) %

\item \textbf{Open-Source Implementation Enabling New Research Directions.} \method offers a flexible and scalable interface that allows researchers to optimize adversarial textures on 3D objects of arbitrary shape by changing only \textbf{one line of code}.  %
This flexibility, combined with the realism of high-fidelity simulation, opens the door to novel research directions in physical adversarial attacks 
(\autoref{sec:experiments}).
\method{}'s code is available at \url{https://github.com/poloclub/undream} and is ready for public release.

\end{enumerate}

\newcommand{\n}{\roundedcell{red!0}{\ding{55}}}
\newcommand{\ygreen}{\roundedcell{green!20}{\ding{51}}}
\newcommand{\y}{\roundedcell{white!20}{\ding{51}}}

\begin{table}[t]
    \caption{\method surpasses existing simulation tools, uniquely standing as the only framework that offers differentiable rendering while simultaneously supporting customizable asset editing in realistic environments.}
    \smallskip
    \setlength{\tabcolsep}{5pt} 
    \centering
    \renewcommand\arraystretch{1.2}
    \small
    \begin{tabular}{@{}l c c c c c@{}}
    \toprule
   \textbf{} & \textbf{\makecell{Code / Tool\\ Available}}& \textbf{\makecell{Edit\\ Assets}} & \textbf{\makecell{Custom\\Assets}} & \textbf{\makecell{Differentiable}} & \textbf{\makecell{Rendering / Tech\\Stack}} \\
    \midrule
    \textbf{\method} (Ours) & \ygreen & \ygreen & \ygreen & \ygreen & \roundedcell{green!20}{Unreal 5.7 (Latest)}\\
    \hdashline[1pt/1pt]
    Carla \cite{dosovitskiy2017carla} & \y & \y & \y & \n & Unreal 4.26 (2020)\\
    AirSim  \cite{shah2017airsim} & \y & \y & \n & \n & Unreal 4.27 (2021)\\
    Carla-Gear \cite{nesti2024carla} & \n & \y & \n & \n & Unreal 4.26 \\
    Carla Drone \cite{meier2024carla} & \y & \n & \n & \n & Unreal 4.26 \\
    SkyScenes \cite{khose2024skyscenes} & \y & \y & \n & \n &  Unreal 4.26\\
    SynDrone \cite{rizzoli2023syndrone} & \y & \y & \n & \n &Unreal 4.26\\
    UrbanScene3D \cite{lin2022capturing} & \y & \y & \n & \n & Unreal 4.27 \\
    \bottomrule
    \end{tabular}
    \label{tab:comparison}
\end{table}

\section{Related Work}

\subsection{Simulation in Adversarial Machine Learning}
High-fidelity simulators have become a cornerstone for developing and testing autonomous systems \cite{wu2020physical, zhang2024visual, nesti2022evaluating} as they bridge adversarial optimization and physical-world evaluation by offering photorealistic graphics, dynamic environments, and configurable sensor suites. 
Simulators such as CARLA~\cite{dosovitskiy2017carla} and Airsim \cite{shah2017airsim}, both of which are built on Unreal Engine \cite{engine2018unreal}, excel at modeling real-world conditions including variable weather, lighting, and camera perspectives making them ideal for generating diverse and scalable datasets~\cite{fonder2019midair, wang2020tartanair, lin2022capturing, rizzoli2023syndrone, khose2024skyscenes, meier2024carladrone} that offer enhanced controllability and scalability compared to real-world benchmarks~\cite{nigam2018ensemble, lyu2020uavid, cai2025vdd}, enabling more systematic evaluation of perception models. However, all of them are non-differentiable.  
\autoref{tab:comparison} shows a detailed comparison of \method{} and existing frameworks, highlighting it as the only framework that simultaneously allows environment customization and differentiation of textures
on the 3D objects in realistic simulation.
While platforms facilitate dataset generation they provide no support for differentiable textures or gradient-based editing in the simulation.
Thus, 
they lack the ability to seamlessly fit in the adversarial optimization pipeline.
As a result, adversarial studies using these frameworks or datasets typically resort to post-hoc superimposition of adversarial textures onto rendered images.
This approach breaks the rendering consistency, causing adversarial objects to appear disharmonious with the scene by failing to account for lighting, shadows, and occlusion \cite{xu2024imperceptible}. 
This limitation not only reduces visual realism, but also severely undermines the transferability and effectiveness of such attacks in real-world scenarios \cite{nesti2022evaluating}.

\subsection{Differentiable Rendering}
Differentiable rendering bridges the gap between traditional computer graphics and gradient-based optimization by enabling gradients to propagate from an image-based loss function back to underlying scene parameters, such as geometry, lighting, and textures~\cite{loper2014opendr, kato2017neural, laine2020modular}.  
This capability has proven especially valuable in adversarial machine learning, as it allows adversarial textures to be optimized directly on 3D objects under varying viewpoints and illumination conditions~\cite{athalye2018synthesizing, zhou2024rauca, li2024flexible, jia2025cca}. 
However, a significant limitation persists: 
differentiable renderers like Mitsuba and Pytorch3D do not support simulation features or complex weather and often require specialized knowledge and manual scene configuration. 
Thus, adversarial researchers prefer using non-differentiable platforms that offer simulation capabilities like Unreal Engine or CARLA instead
\cite{hull2024renderbender}.

These limitations have created a long-standing divide: researchers must choose either \textbf{high-fidelity, non-differentiable simulation} that requires gradient approximation techniques~\cite{xu2024imperceptible}, or \textbf{differentiable renderers with restricted scene dynamics} that constrain the creation of photorealistic simulations and animations and demand specialized expertise and manual scene configuration~\cite{laine2020modular, kato2017neural, loper2014opendr}. Our work closes this gap by offering a framework that bridges differentiable optimization to high-fidelity 3D environments.

\section{Method}

\label{sec:method}

One of the core contributions of \method is its ability to bridge a longstanding gap between high-fidelity simulation and differentiable rendering --- two critical components required for end-to-end optimization in simulation-based environments.
Simulation platforms commonly used in adversarial machine learning research provide highly realistic visual outputs and complex scene dynamics, but lack native support for gradient-based optimization. 
In contrast, differentiable rendering frameworks enable precise, gradient-based optimization but are limited in their ability to generate complex, photorealistic environments with dynamic elements.

\noindent \textbf{Selection of simulation platform.} While there is a range of platforms to choose from when it comes to picking the simulations, most of the widely used simulation platforms are built on Unreal Engine \cite{dosovitskiy2017carla, meier2024carladrone, shah2017airsim} making it a natural choice. We implement \method using Unreal Engine 5 --- instead of the Unreal Engine 4 used in the previously mentioned platforms --- as it introduces new light scattering techniques such as Nanite and Lumen that improve photorealism in rendering \cite{epicgamesLumenGlobal}. 

\noindent \textbf{Selection of differentiable renderer.} Mitsuba is a popular differentiable renderer used in adversarial attacks \cite{hull2024renderbender}. We use Mitsuba in \method implementation as it a has Python API which allows for integration with existing adversarial ML implementations. Mitsuba, due to its differentiable ray tracing method, has increased photorealism compared to alternatives such as PyTorch3D \cite{ravi2020accelerating}.

 \method unifies the strengths of these two paradigms by combining the visual realism and scene richness of Unreal Engine with the differentiability and optimization capabilities of Mitsuba. This enables gradient-based adversarial optimization directly within photorealistic simulations. Moreover, \method is implemented entirely in Python, ensuring seamless compatibility with widely adopted adversarial machine learning frameworks, particularly those built on PyTorch.

\subsection{Setup} 
\subsubsection{Forward Rendering: Sequences in Unreal Engine}
A LevelSequence is Unreal Engine’s native tool for constructing cinematics and is used to capture dynamic scenes with precise temporal and spatial coherence. Users may either import an existing LevelSequence into their Unreal Engine project or create a new sequence from scratch. When creating a new sequence, converting key scene elements such as cameras and dynamic objects into spawnable assets ensures accurate tracking of object motion and reliable reproduction of scenes across frames. 

\begin{figure}[t]
    \centering
    \includegraphics[width=
    \textwidth]{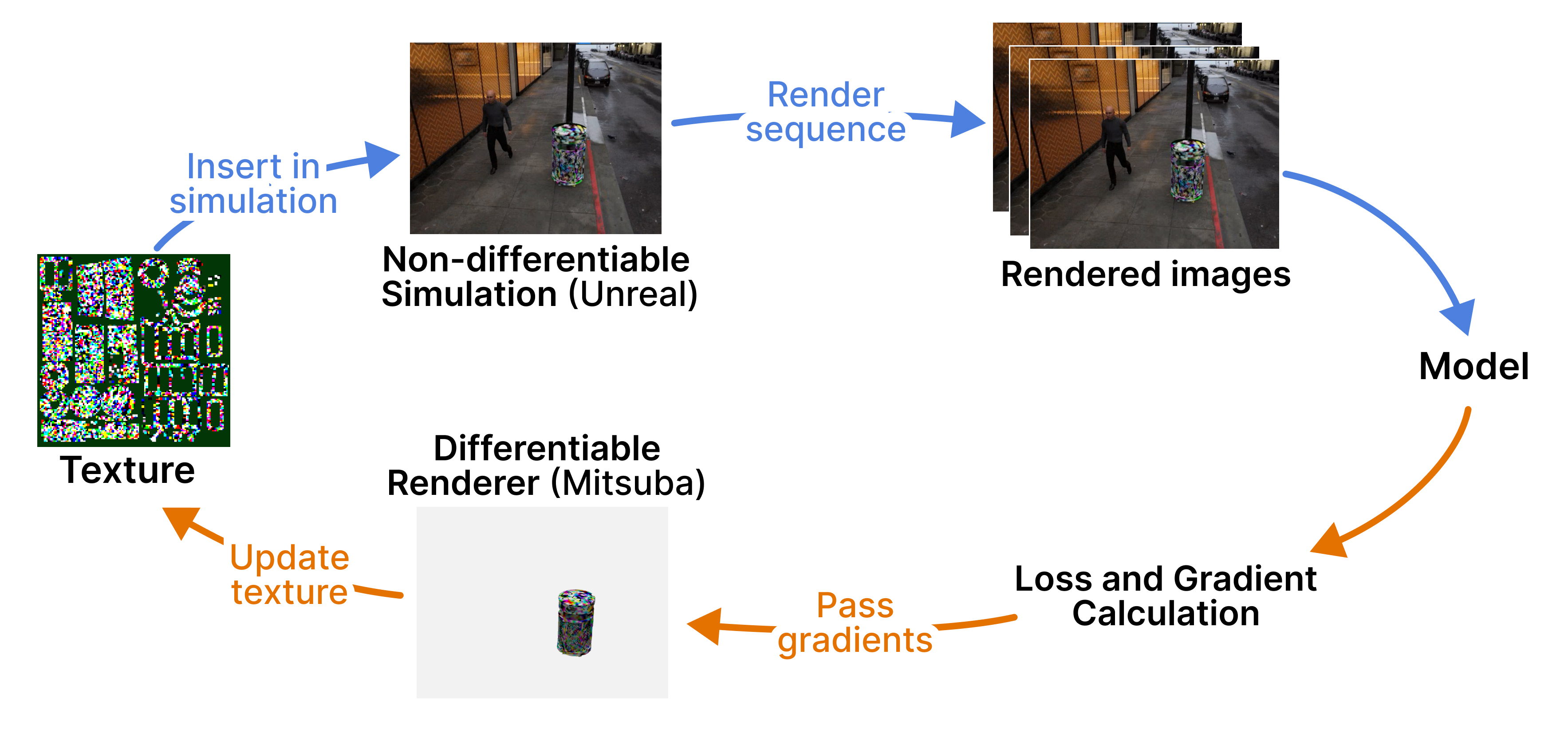} 
    \caption{The end-to-end \method optimization pipeline. In each iteration, the adversarial texture is applied to the object in simulation, model predictions are computed, loss is calculated and the gradients are propagated back through the object in the simulation equivalent XML scene to optimize the adversarial texture.}
    \label{fig:pipeline}
\end{figure}

\subsubsection{Inverse Rendering: Scene Transformation.}
\label{sec:inverse-rendering}
Transforming scenes from Unreal Engine to Mitsuba involves reconciling differences across several axes.  \method converts \textit{left-handed Z-up} coordinate system, with \textit{Y-facing} cameras, \textit{Euler} rotations, and \textit{meter} units from Unreal Engine into \textit{right-handed Y-up} coordinate system, with \textit{Z-facing} cameras, \textit{axis-based} rotations, and \textit{centimeter} units in Mitsuba, preserving consistent object–camera geometry across renderings.
This transformation process is fully automated and can be executed via the \textit{``initialize.py''} script provided in the project repository.
The transformation pipeline involves the following key steps:

\noindent \textbf{Extract camera and object poses.} 
Retrieve information from the LevelSequence including the 3D positions and orientations of adversarial object, camera, and any other object we would like to track across the frames.

\noindent \textbf{Apply scale and coordinate conversion.} 
Transform retrieved location and rotation from the LevelSequence from the Unreal Engine coordinate system to the Mitsuba coordinate system and convert the relevant units. This includes adjustments to axes, handedness, and rotation conventions.

\noindent \textbf{Generate Mitsuba XML scenes.} 
For each frame in the sequence, using the transformed object positions, orientations, and camera parameters create XML files that replicate the relationship between the camera and adversarial object such that the object appears exactly as it does in the simulation. 

\noindent \textbf{Add adversarial texture to XML.} 
Restore the desired adversarial texture in the XML files in preparation for gradient-based optimization of adversarial texture.

Another intermediate step is replacing the object's texture with a plain white texture in the XML scenes, and save the resulting rendering. This can help us verify XML rendering. 
Further details about the transformation can be found in \autoref{label: appendix-unreal-mitsuba transfer}

\subsection{Run adversarial attack}
\label{sec:attack-pipeline}

While existing attack libraries support a range of gradient-based attacks, their implementations present several limitations in the context of differentiable simulation \cite{nicolae2018adversarial}: 
(1) they do not support integration with differentiable renderers, 
(2) they lack mechanisms for inserting updated textures into a simulation at each optimization step, and 
(3) they do not allow dynamically updating input images based on the output of new renderings after each iteration.
\method enables smooth adaptation of attacks as shown in \autoref{fig:pipeline} with the Algorithm \autoref{algo:attack}.

The attack implementation in \method builds upon existing adversarial attacks available in widely used attack libraries such as ART \cite{nicolae2018adversarial}, with  modifications to enable integration with the differentiable rendering process, as detailed in the bridging step of Algorithm~\ref{algo:attack}. 
Our pipeline iteratively renders scenes, computes gradients for differentiable rendering, and updates textures within simulation.
Integrating new attacks into the \method pipeline requires modifying only \textbf{a single line of code} --- the line responsible for updating the adversarial texture using the computed gradient in the differentiable renderer as shown in Line \ref{algo-step:diff renderer}. By altering this line, users can easily switch between different attack methods supported by existing attack libraries while retaining full compatibility with the differentiable simulation workflow.

\renewcommand{\algorithmicrequire}{\textbf{Input:}}

\begin{algorithm}
\caption{\method Attack Pipeline Generates Adversarial Texture by \textit{Bridging} Unreal Simulation and Differentiable Rendering}
\begin{algorithmic}[1]
\Require input scene $S$ (LevelSequence), victim model  $M$, attacker-chosen output $y$, initial texture  $T_i$, attack iterations $j$ (equal to number of rendering jobs) 
\Ensure Adversarial texture $T_{adv}$
\State \textbf{Initialize:} $T_{adv}  \leftarrow  T_{i}$
\For {attack iteration $j$}
    \State $x_{i-n}$ $\leftarrow$ Images from input scene $S$ 
    \Procedure{Rendering job finish callback}{} 
       \State Get model predictions  $pred \leftarrow M(x)$
       \State Find $L \leftarrow Loss(y,pred)$
        \State Calculate gradients $g \leftarrow \nabla L$
        \State  \colorbox{lightyellow}{\textbf{Pass gradients to differentiable renderer} } \Comment{Bridging step}
        \State \textit{Using differentiable renderer:}  Use gradients to update texture 
        \label{algo-step:diff renderer}
      \State  Save updated $T_{adv}$
      \State  Update new $T_{adv}$ as object texture in Unreal Engine
   \EndProcedure
\EndFor
\end{algorithmic}
\label{algo:attack}
\end{algorithm}

\section{Experiments}
\label{sec:experiments}
We evaluate the capabilities of \method along two primary axes: the level of control afforded within the simulation environment, and the flexibility and effectiveness of adversarial attacks implemented through the framework. These two dimensions highlight \method's utility both as a research tool for simulation-based adversarial machine learning.
An overview of the available controls spanning both simulation and optimization components is presented in \autoref{tab:tool}. This includes parameters such as scenes, objects, lighting, models, tasks and algorithms.

\subsection{Object}
\begin{figure}[t]
\centering
\begin{subfigure}[b]{0.43\textwidth}
  \centering
  \includegraphics[width=\textwidth]{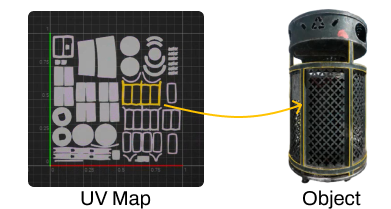}
  \caption{}
  \label{fig:uv}
\end{subfigure}
\hfill
\begin{subfigure}[b]{0.48\textwidth}
  \centering
  \includegraphics[width=\textwidth]{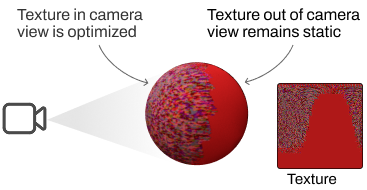}
  \caption{}
  \label{fig:optimize-view}
\end{subfigure}
\caption{\textbf{(a)} UV layout and object mesh in the UV editor display in Unreal Engine 5. Each segment from the UV map corresponds to a part on the object. \textbf{(b)} Optimized object (sphere) shown from the side. Only the part of the texture that is visible in the camera view is optimized.}
\label{fig:uv-camera-views}
\end{figure}

\begin{figure}[b]
    \centering
    \includegraphics[width=
    \textwidth]{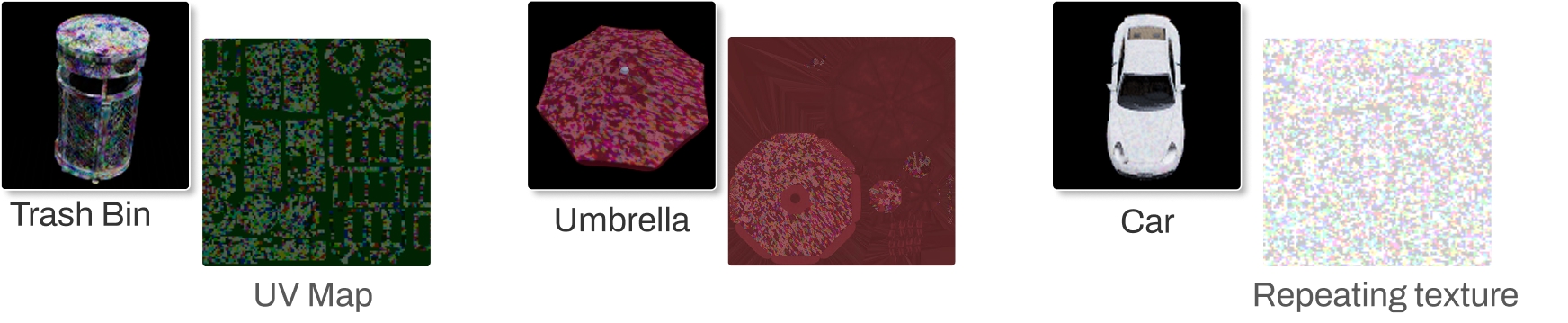} 
    \caption{\method enables optimization of 3D objects of arbitrary shapes. For large objects (e.g., car), Unreal allows the texture to repeat, thus covering the whole object.}
    \label{fig:objects}
\end{figure}

One of the main contributions of \method is it enables researchers to optimize adversarial textures for any arbitrary 3D object in high-fidelity simulations. This will open many new research directions as the current optimization pipeline uses 2D approximation techniques when dealing with photorealistic simulations \cite{xu2024imperceptible}, thus limiting the adversarial objects possible to only 2D rectangular shapes.
Objects in the simulation utilize a technique known as UV mapping to project a 2D texture onto their 3D surface. This mapping allows us to determine how each pixel in the texture corresponds to a specific point on the object’s surface, facilitating precise gradient-based updates during optimization. An example of the UV map editor in Unreal Engine is shown in ~\autoref{fig:uv}. 
Upon completion of the optimization process, we observe that only the regions of the texture directly visible to the camera are significantly modified, as illustrated in ~\autoref{fig:optimize-view}.
~\autoref{fig:objects} presents several examples of objects used in our experiments, along with their optimized adversarial textures.

\begin{figure}[t]
    \centering
    \includegraphics[width=
    \textwidth]{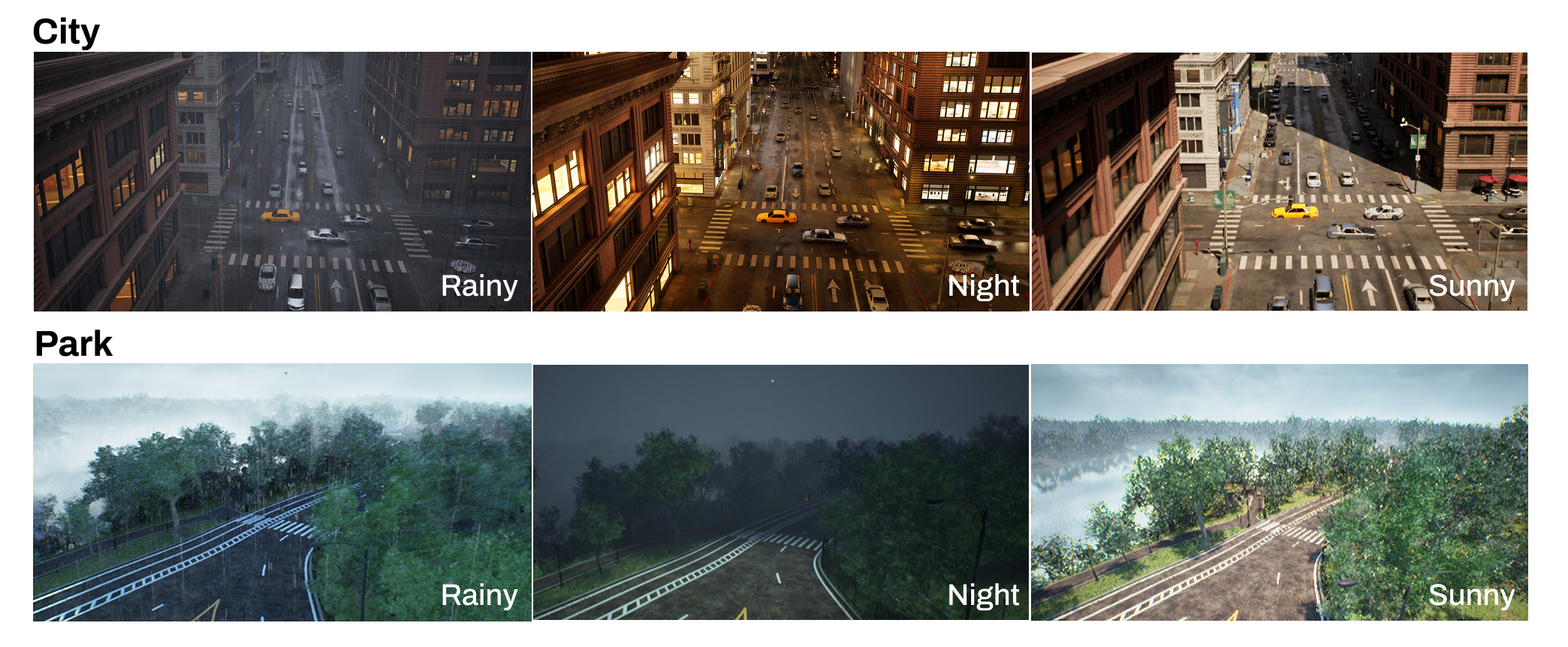} 
    \caption{The same scene rendered under different lighting and weather conditions using \method demonstrating its ability to replicate environmental factors. A city scene (top) and a park scene (bottom) are rendered in rainy, night and sunny environments.}
\label{fig:environment}
\end{figure}

\subsection{Environment and Lighting}
Environmental and lighting conditions play a critical role in adversarial machine learning, particularly in the context of physically realizable attacks. 
Variations in lighting, shadows, time of day, or weather can significantly influence the effectiveness and transferability of adversarial attacks \cite{wei2024physical}.
However, existing datasets and simulation pipelines typically offer limited support for systematically varying environmental parameters. 
Even when present, they are often not available as controllable inputs for experimentation, but preset variables \cite{nesti2024carla}.
In contrast, \method provides fine-grained control over lighting and environmental conditions within the simulation, enabling researchers to evaluate the robustness of adversarial attacks across diverse settings. ~\autoref{fig:environment} demonstrates examples of the same scene rendered under different lighting and weather conditions using \method.

\begin{table}[b]
    \centering
    \caption{Overview of configurable elements supported by \method, including simulation components and adversarial attack settings.}
    \smallskip
    \setlength{\tabcolsep}{4pt} 
    \renewcommand\arraystretch{1.1}
    \begin{tabular}{l l l l}
    \toprule
        \multicolumn{2}{c}{\textbf{Element}} & \multicolumn{1}{l}{\textbf{Possibilities}} & \multicolumn{1}{l}{\textbf{Demonstrated}} 
        \\
        \midrule
        \multirow{3}*{Simulation}  & Objects & \textbf{Any} object (.OBJ) & Sphere, bin, car, umbrella \\
                                  & Scenes & \textbf{Any} scene & Park, city \\
                                  & Lighting & \textbf{Any} lighting or weather variation & Sunny, cloudy, dark, rainy \\
        \hline
        \multirow{3}*{Attacks}  & Models & \textbf{Any} model from HuggingFace & Detr-resnet-50 \\
                                  & Tasks & Can be expanded to \textbf{any} tasks & Object detection, classification \\
                                  & Algorithms & Can be expanded to \textbf{any} attack & PGD, Auto-PGD \\
        \bottomrule
    \end{tabular}

    \label{tab:tool}
\end{table}

\subsection{Attacks and Tasks}
\method enables seamless experimentation with a wide range of adversarial attacks and tasks within a high-fidelity simulation environment.
As described in Algorithm~\ref{algo:attack}, the structure of the attack implementation in \method mirrors that of popular adversarial attack libraries. 
This ensures that attacks can be easily adapted and integrated into the simulation with minimal code modifications.
As examples to demonstrate successful attack execution, \autoref{tab:attack} shows the results for the PGD and Auto-PGD attacks for the tasks of classification and detection on detr-resnet-50 for the sequences provided in the code repository; 
reducing the classification accuracy from 100\% to 9.5\%, and detection mAP from 100\% to 6.9\% a  significant drop commonly seen in unbounded attacks \cite{xu2024imperceptible}.
Our code repository includes implementations for attacking a variety of models (e.g., detr-resnet-50, YOLOv8, YOLOv11).

\begin{table}[h!]
    \centering
    \renewcommand\arraystretch{1}
    \caption{Attack results for attacks in \method with \textit{classification} accuracy and \textit{detection} mAP (mean average precision).}
    \smallskip
    \setlength{\tabcolsep}{4pt} 
    \label{tab:attack}
    \begin{tabular}{l c c c c c}
    \toprule
    \textbf{Attack} & \textbf{Budget} & \multicolumn{2}{c}{\textbf{Benign}} & \multicolumn{2}{c}{\textbf{Adversarial}} \\
    \cmidrule(lr){3-4} \cmidrule(lr){5-6}
    &  & Accuracy (\%)  & mAP & Accuracy (\%) & mAP \\
    \midrule
    \multirow{4}{*}{PGD}  & 50 & \multirow{4}{*}{100} & \multirow{4}{*}{100} & 33.0 & 45.3 \\
    & 100 & & & 31.6 & 45.0\\
    & 200 & & & 23.8 & 29.5\\
    & 255 & & & 9.5 & 14.3\\
    \arrayrulecolor{gray!60}\hline
    \multirow{4}{*}{Auto-PGD} & 50 & \multirow{4}{*}{100} & \multirow{4}{*}{100}  & 33.0 & 40.2\\ 
    & 100 & & & 28.3  & 32.8 \\
    & 200 & & & 20.0 & 19.8\\
    & 255 & & & 9.5 & 15.8\\
    \bottomrule
    \end{tabular}
\end{table}

\section{Conclusions and Future Work}

We presented
\method, the first software framework bridging differentiable rendering and photorealistic simulation for end-to-end adversarial attacks.
\method offers automatic, precise 3D transformation enabling faithful threat modeling and surpasses existing simulation tools, uniquely standing as the only framework that offers differentiable rendering while simultaneously supporting customizable asset editing in realistic environments.
Its open-source implementation enables new research directions.
\method currently supports a wide variety of scenes, sequences, camera angles, object trajectories, 3D models, and lighting configurations. 
With \method{}'s open-source implementation, researchers and developers may easily
extend \method{} to support more simulation assets, attacks (\autoref{sec:attack-pipeline}), and even defenses,
and test them all under realistic, high-fidelity conditions.
We are excited by the possibilities enabled by \method in bridging the gap between photorealistic evaluations and physical-world deployment, further strengthening research on adversarial robustness.

\bibliographystyle{splncs04}
\bibliography{bibliography}

\appendix
\section{Transformation details}
\label{appendix:transform}
Transformation of a scene from Unreal Engine to Mitsuba is the first fundamental step in bridging simulation with a differentiable renderer. The difference between these platforms are detailed in \autoref{tab:diff-unreal-mitsuba}. 
\label{label: appendix-unreal-mitsuba transfer}
While scale conversion is relatively easy, the steps to tackle coordinate and rotation conversion are expanded on here:

\paragraph{World coordinate:}
To convert Left handed Z up to a Right handed Y up system, we must swap values on the Y and Z axes. Keep in mind this also affects rotation, as the rotation for Mitsuba is defined by coordinate axes. 

\paragraph{Rotation:}
Unless the pitch (rotation around Y) in Unreal Engine is a $\pm$ 90 degrees, we can simply rotate the camera and object at the origin before translation. 
However, if the Y rotation is $\pm$ 90 degrees, the object is subject to Gimbal Lock, a phenomenon in which the object loses one of its degrees of freedom, thus rotations around Z axis, may not actually apply. 

\begin{table}[h]
    \centering
    \caption{Comparison of coordinate systems and transformation conventions between Unreal Engine and Mitsuba.}
    \setlength{\tabcolsep}{10pt} \renewcommand\arraystretch{1.2}
    \begin{tabular}{l l l}
    \toprule
        Parameter & Unreal Engine & Mitsuba\\
         \midrule
        World Coordinate & Left handed Z up & Right handed Y up\\
        \multirow{2}{*}{Camera orientation} & Facing Y & Facing Z \\
        & Z up& Y up\\
        Rotation & XYZ Euler & Coordinate axes\\
        Scale & meters & centimeters\\
        \bottomrule
    \end{tabular}
    
    \label{tab:diff-unreal-mitsuba}
\end{table} 
\end{document}